\documentclass[conference, a4paper]{IEEEtran}
\IEEEoverridecommandlockouts

\ifCLASSOPTIONcompsoc
  \usepackage[nocompress]{cite}
\else
  \usepackage{cite}
\fi

\ifCLASSINFOpdf
\usepackage[pdftex]{graphicx}
\graphicspath{{../pdf/}{../jpeg/}}
\DeclareGraphicsExtensions{.pdf,.jpeg,.png}
\else

\usepackage[dvips]{graphicx}\graphicspath{{../eps/}}
\DeclareGraphicsExtensions{.eps}
\fi
\usepackage{amsmath}
\usepackage{algorithmic}
\usepackage{array}
\ifCLASSOPTIONcompsoc
 \usepackage[caption=false,font=footnotesize,labelfont=sf,textfont=sf]{subfig}
\else
\usepackage[caption=false,font=footnotesize]{subfig}
\fi

\usepackage{fixltx2e}
\usepackage{dblfloatfix}
\usepackage{url}
\usepackage[table,xcdraw]{xcolor}
\usepackage{multirow}
\usepackage{tabularx}

\def\BibTeX{{\rm B\kern-.05em{\sc i\kern-.025em b}\kern-.08em
    T\kern-.1667em\lower.7ex\hbox{E}\kern-.125emX}}

\usepackage{tikz}

\newcommand\copyrighttext{%
  \footnotesize \textcopyright \the\year{} IEEE. Personal use of this material is permitted. Permission from IEEE must be obtained for all other uses, including reprinting/republishing this material for advertising or promotional purposes, collecting new collected works for resale or redistribution to servers or lists, or reuse of any copyrighted component of this work in other works.}

\newcommand\copyrightnotice{%
\begin{tikzpicture}[remember picture,overlay]
\node[anchor=south,yshift=10pt] at (current page.south) {\fbox{\parbox{\dimexpr0.75\textwidth-\fboxsep-\fboxrule\relax}{\copyrighttext}}};
\end{tikzpicture}%
}

\begin{document}

\title{Congestion or No Congestion: Packet Loss Identification and Prediction Using Machine Learning}

\author{\IEEEauthorblockN{Inayat Ali}
\IEEEauthorblockA{Network Research Division\\
ETRI\\
Daejeon, Gajeong-ro 218\\
Email: inayat@etri.re.kr}
\and

\IEEEauthorblockN{Seungwoo Hong}
\IEEEauthorblockA{Network Research Division\\
ETRI\\
Daejeon, Gajeong-ro 218\\
Email: swhong@etri.re.kr}
\and
\IEEEauthorblockN{Taesik Cheung}
\IEEEauthorblockA{Network Research Division\\
ETRI\\
Daejeon, Gajeong-ro 218\\
Email: 	cts@etri.re.kr}}

\maketitle
\copyrightnotice
\begin{abstract}
Packet losses in the network significantly impact network performance. Most TCP variants reduce the transmission rate when detecting packet losses, assuming network congestion, resulting in lower throughput and affecting bandwidth-intensive applications like immersive applications. However, not all packet losses are due to congestion; some occur due to wireless link issues, which we refer to as non-congestive packet losses. In today's hybrid Internet, packets of a single flow may traverse wired and wireless segments of a network to reach their destination. TCP should not react to non-congestive packet losses the same way as it does to congestive losses. However, TCP currently can not differentiate between these types of packet losses and lowers its transmission rate irrespective of packet loss type, resulting in lower throughput for wireless clients. To address this challenge, we use machine learning techniques to distinguish between these types of packet losses at end hosts, utilizing easily available features at the host. Our results demonstrate that Random Forest and K-Nearest Neighbor classifiers perform better in predicting the type of packet loss, offering a promising solution to enhance network performance.
\end{abstract}


\IEEEpeerreviewmaketitle

\section{Introduction}
Internet users are growing every year and most users, over about 96.3\% use their smartphones to access the Internet. Mobile devices contribute to 60\% of global web traffic \footnote{\url{https://datareportal.com/global-digital-overview}}. This shows the importance of wireless technologies which are developing at a greater pace. However, the current TCP performs less efficiently in a wireless environment compared to a wired environment due to various reasons. One of the main reasons is the higher packet loss rate due to wireless characteristics such as signal interference, fading, mobility, and physical obstructions. We call these types of packet losses \textit{non-congestive packet losses} while we call packet losses due to congestion as \textit{congestive packet losses}. TCP interprets these losses as a sign of network congestion, leading to unnecessary retransmissions and reduced transmission rates that affect network performance. There is no direct way to identify these non-congestive packet losses at the end host and react accordingly. \\

Few TCP variants are optimized to perform better in wireless environments by estimating the congestion based on other parameters instead of using packet loss as a measure of congestion. For example, TCP Westwood \cite{ref1} continuously estimates the available bandwidth by measuring the rate of returning acknowledgments (ACKs). When packet loss occurs, TCP Westwood adjusts the congestion window (cWnd) and the slow start threshold (SsThresh) based on the estimated bandwidth. However, efficiently estimating bandwidth itself is challenging. Similarly, TCP Veno \cite{ref2} monitors the congestion level of the network by measuring the difference between the expected and actual round-trip times (RTT). This measurement helps estimate the queueing delay. When packet loss is detected, TCP Veno uses the queueing delay to determine whether the loss is due to congestion. If the queueing delay is small, the loss is likely due to wireless errors; if the queueing delay is large, the loss is likely due to congestion. However, these mechanisms still do not perform adequately in predicting non-congestive packet losses. 

Machine learning has been extensively adopted in computer networking to address a wide range of issues, including threat detection, traffic pattern prediction, and network resource management \cite{ref3}. Recently, ML techniques have also been utilized to manage congestion by predicting retransmissions \cite{ref4} and using reinforcement learning to adjust congestion windows \cite{ref5}. Additionally, recent efforts have focused on using machine learning for packet loss prediction, yielding significant results \cite{ref6}\cite{ref7}. However, these studies do not directly predict and differentiate between types of packet loss. 
In this work, we leverage various machine learning models to not only predict packet losses but also distinguish between congestive and non-congestive packet losses. When a packet loss is identified as non-congestive packet loss, the client should not drop its congestion window as this loss is not due to congestion in the network but a random loss due to temporary wireless link conditions. This will ensure relatively high throughput in a wireless environment.    

Subsequent sections of this paper are organized as follows: Section 2 provides a brief overview of the literature on loss prediction and congestion management. In Section 3, we discuss the problem statement in detail. We explain the dataset in section 4. Section 5 presents the methodology used in this study. We discuss the results in section 6. Finally, we conclude the paper in Section 7.

\section{Related Work}

Some TCP variants optimized for wireless environments manage congestion using parameters other than packet loss, such as delay, jitter, and bandwidth estimation \cite{ref1}\cite{ref2}. These rule-based variants often fail to accurately estimate network conditions, resulting in limited improvement in wireless environments. Furthermore, their performance may degrade in hybrid network paths that include both wired and wireless links \cite{ref8}. The authors in \cite{ref6} propose a mathematical model for loss prediction in real-time audio based on delay variation, combining them with weights according to their importance. However, this approach only achieved an accuracy of 60\% to 90\%, which is significantly lower than the accuracy achieved by advanced machine learning models.

Giannakou \cite{ref4} proposed a machine-learning tool based on Random Forest for predicting packet retransmission in science flows. This model considers path properties like RTT and host parameters, including the congestion window, to predict the number of retransmitted packets in each network flow, achieving a prediction accuracy of 97\% to 99\%. Similarly, to ensure Quality of Experience (QoE) in real-time communication (RTC), the authors in \cite{ref7} propose a Random Forest classifier-based model that accurately predicts packet loss for RTP streaming traffic. They tested several machine learning models using data from popular RTC applications such as Cisco Webex and Jitsi Meet. Although these methods effectively predict packet loss, they do not distinguish between congestive and non-congestive packet loss.

Additionally, Naive Bayes and Support Vector Machine (SVM) methods have been applied to predict congestion at network routers using packet statistics, such as the number of packets in the queue, remaining space in the queue, and the current queue average \cite{ref9}. While Naive Bayes outperforms SVM in predicting network congestion, this approach faces practical challenges. Vendor-specific routers and switches often do not support third-party daemons and implementing this mechanism on every router between the server and client is difficult. Moreover, the Explicit Congestion Notification (ECN) \cite{ref12} \cite{ref13} mechanisms can explicitly notify congestion at router queues, making this solution less effective in practice.

\section{Problem Statement}
\begin{figure*}[h]
\centering
\subfloat[Throughput]{\includegraphics[width=2.48in]{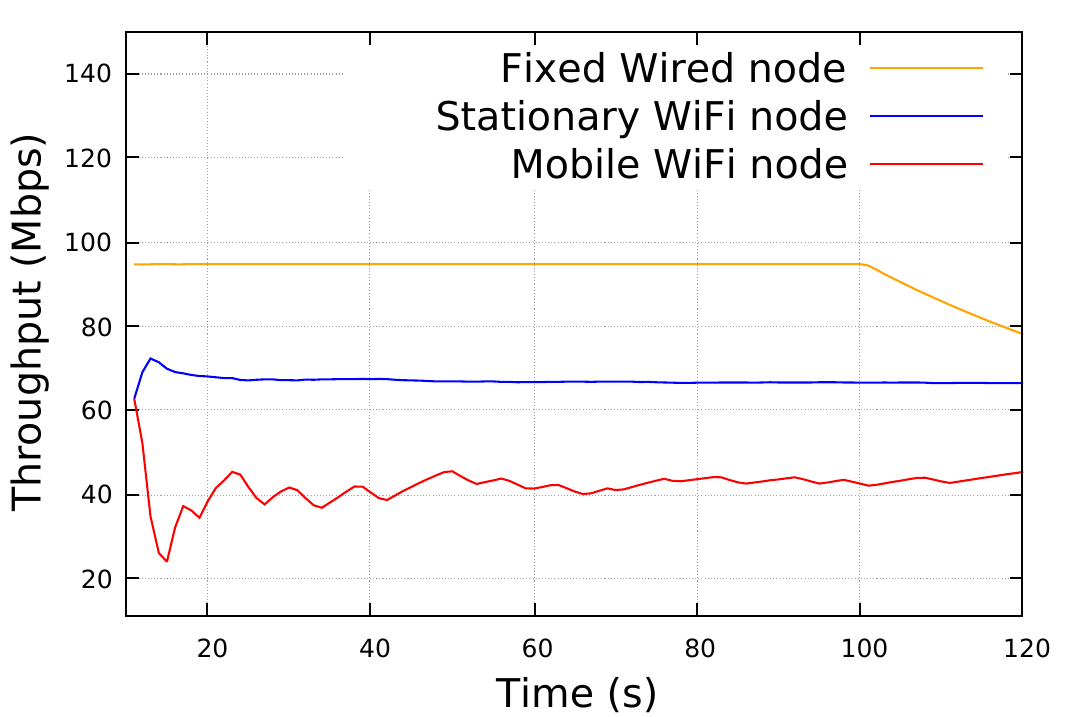}%
\label{fig_thro}}
\hfil
\subfloat[Congestion Window]{\includegraphics[width=2.6in]{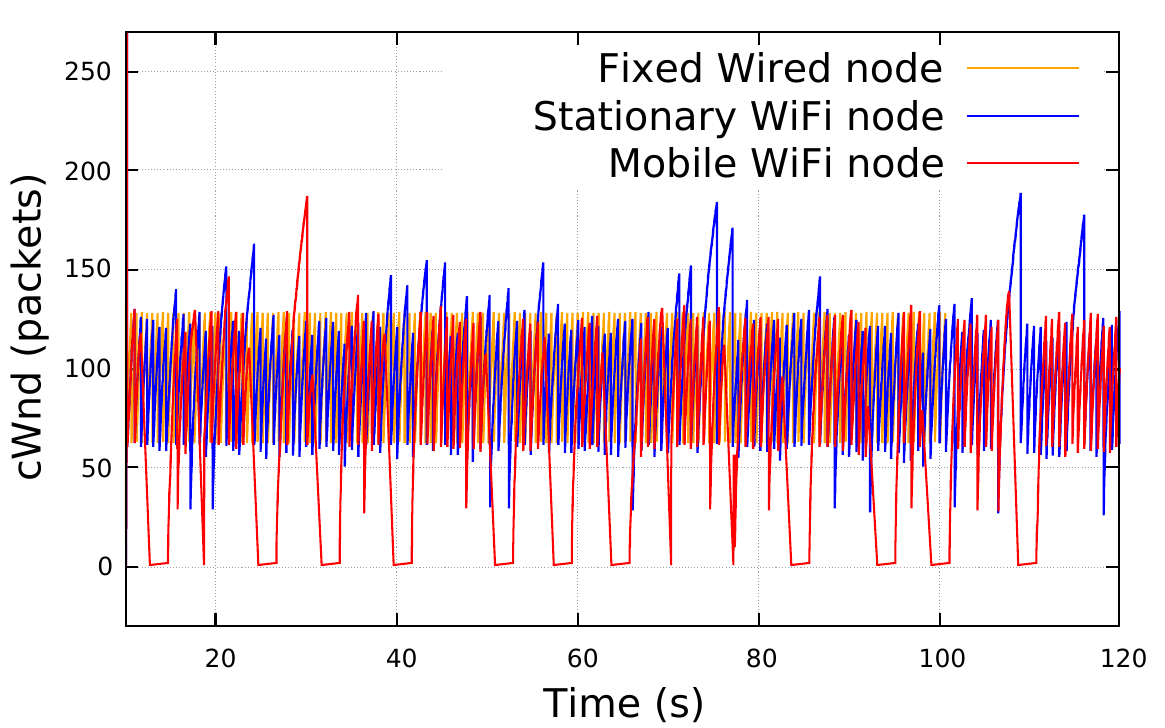}%
\label{fig_cwnd}}
\caption{Comaprison of stationary/mobile wireless client with wired client}
\label{fig_sim}
\end{figure*}

Mobile users access the Internet through wireless technologies like WiFi or LTE/5G, accounting for over 60\% of global web traffic. However, standard TCP performs poorly in wireless environments due to the inherent physical constraints of wireless links. Issues such as wireless channel interference, fading, and mobility lead to additional packet losses, triggering more retransmissions and reducing the sender's transmission rate (cWnd reduction), which significantly affects throughput, as illustrated in Figure \ref{fig_thro}.

In Figure \ref{fig_thro}, three nodes are connected to the same AP/router: one via a wired network and two via WiFi. Among the WiFi-connected nodes, one is stationary while the other is mobile, causing packet drops at the wireless link due to lower signal strength. All three nodes retrieve data from the same remote server simultaneously. The results show that the stationary WiFi node achieves lower throughput compared to the wired node, even without packet loss at the wireless channel. This is due to the physical constraints of the wireless channel. The mobile node experiences much lower throughput due to packet losses at the wireless channel. Although these losses are not caused by congestion, the server mistakenly treats them as such, reducing its cWnd to alleviate perceived congestion, resulting in lower throughput. Figure \ref{fig_cwnd} shows the cWnd behavior for the three flows, with the mobile host's TCP flow exhibiting more variation due to non-congestive packet losses causing frequent cWnd reduction.

%\begin{figure}[!t]
%\centering
%\includegraphics[width=2.5in]{figures/throughput.pdf}
%\caption{Throughput}
%\label{fig_thro}
%\end{figure}

\section{Dataset and Loss characterization}

In this section, we will provide an overview of our dataset and the types and number of packet losses that occur. The dataset was collected from a simulation environment using ns-3 \cite{ref10}, an open-source simulator for internet systems. ns-3 closely mimics a real network environment, simulating the entire TCP/IP stack and generating real packets that are sent over simulated links.

Our setup involves a single long TCP flow between a fixed server and a mobile client. The server is connected to a wired network, while the remote mobile client connects to a WiFi access point (AP) a few hops away from the server. The RTT between the server and the mobile client varies from 1 to 100 ms. During the simulation, the server transmits a large file to the mobile client using TCP. Data collection occurs on the server side in \textit{pcap} format using the Wireshark packet sniffer. \\

\subsection{Packet loss and feature selection} Once the server receives the ACK for the first packet, the data collector records crucial features for model training. In order to select the best feature for our machine learning model, we thoroughly investigated the relationship between packet drop and the flow metrics. These features include timestamp, packet size, RTT, average RTT, jitter, and congestion window size. The features are selected after a thorough study of loss packets and we find a correlation of these features and discuss it in a later section. The flow statistics are summarized in Table \ref{table-1}.

In Table \ref{table-1}, \textit{qDrop} refers to queue drops, which are packets dropped at intermediate routers' queues due to congestion. In contrast, \textit{wDrop} denotes wireless drops, which are packets lost due to wireless channel issues, representing non-congestive packet drops. There are a total of 1.66\% of packets that are lost among which $3990$ are lost due to congestion and $3546$ packets are lost due to wireless channel issues. 

\begin{table}[ht]
\centering
\caption{Summary of packets in the flow}
\begin{tabular}{|c|c|c|}
\hline
\rowcolor[HTML]{C0C0C0} 
\textbf{Description}                     & \textbf{Quantity} & \textbf{Percentage (\%)} \\ \hline
Total packets                            & 451352            & -                        \\ \hline
Total drop packets                       & 7536              & 1.66                     \\ \hline
\rowcolor[HTML]{E6E6E6} 
Congestive drop (\textit{qDrop})         & 3990              & 0.8                      \\ \hline
Non-congestive drop (\textit{wDrop})     & 3546              & 0.78                     \\ \hline
\end{tabular}
\label{table-1}
\end{table}

\section{Proposed Methodology}
In this section, We present different ML models to classify packet loss types. We investigated five different ML models including Random Forest (RF), K-Nearest Neighbor (KNN), Gradient Boosting (GB), Logistic Regression (LR), and Decision Tree (DT) classifier. We follow the same pipeline for all the models: train the model, hyperparameter tuning, and performance evaluation. For validation, we divide the data into 80\% for training and 20\% for testing.

Due to the high imbalance in our dataset, metrics like precision and accuracy are skewed by the predominance of the majority class. While the F1-score provides a balanced measure of overall performance, it is still influenced by precision. Therefore, we have selected recall as our primary performance metric and optimization goal. Recall, defined as the ratio of true positive samples to the total number of actual positive samples, more accurately reflects the model's effectiveness in identifying positive instances in an imbalanced dataset.\\
Note that our classifier is designed to be integrated into the end host TCP protocol, enabling it to respond appropriately when a congestive packet loss is detected while ignoring non-congestive packet losses.

\subsection{Machine learning models}
First, we considered Random forest, gradient boosting, and decision tree classifier as these tree-based algorithms work well with traffic data \cite{ref11}. A Decision Tree classifier is a straightforward yet effective machine learning algorithm that works by recursively partitioning the data into subsets based on the values of input features. Each node in the tree represents a decision point, where the dataset is split according to a feature that best separates the classes. The process continues until the terminal nodes, or leaves, are reached, each representing a class label. This hierarchical structure makes Decision Trees easy to interpret and visualize. Random Forest is an ensemble learning method that enhances the performance of individual Decision Trees by constructing a multitude of them and aggregating their results. Each tree is built from a random subset of the data and features, introducing diversity and reducing the risk of overfitting. The final prediction is determined by averaging the predictions of all trees (for regression tasks) or by majority voting (for classification tasks). This method leverages the strength of multiple models to achieve better generalization and robustness. Gradient Boosting is an advanced ensemble technique that builds models sequentially, with each new model correcting the errors of its predecessor. This approach focuses on instances that are hard to predict, incrementally improving the overall model accuracy. \\
The K-Nearest Neighbor classifier is an instance-based learning algorithm that classifies a data point based on the majority class among its k-nearest neighbors in the feature space. It operates on the principle that similar instances are close to each other. The algorithm calculates the distance between the input data point and all points in the training set, selects the k closest ones, and assigns the most common class among them. KNN is simple to implement and often effective, especially for low-dimensional data. Logistic Regression is a statistical model used for binary or multiclass classification tasks. It estimates the probability that a given input belongs to a particular class using a logistic function. The algorithm models the relationship between the input features and the probability of the class label by fitting a linear combination of the features. This probability is then transformed using the logistic (sigmoid) function to produce a value between 0 and 1.

\begin{table*}[h]
\centering
\caption{Machine Learning Models Performance Comparison}
\begin{tabular}{|l|p{1.5cm}|c|c|c|c|c|c|c|c|c|}
\hline
\textbf{Machine Learning Models} & \textbf{Packet Loss Type} & \textbf{Recall} & \textbf{F1-Score} & \multicolumn{2}{c|}{\textbf{Support}} & \textbf{Macro avg F1} & \multicolumn{3}{c|}{\textbf{Confusion Matrix}} \\ \cline{5-6} \cline{8-10}
 &  &  &  & \textbf{Actual} & \textbf{Pred} &  & \textbf{qDrop} & \textbf{wDrop} & \textbf{unDrop} \\ \hline

\textbf{Random Forest Classifier} & qDrop & 0.94 & 0.97 & 1233 & 1158 & \multirow{2}{*}{0.88} & 1158 & 0 & 0 \\ \cline{2-6} \cline{8-10}
& wDrop & 0.62 & 0.67 & 251 & 212 &  & 53 & 156 & 3 \\ \hline

\textbf{K-Neighbor Classifier} & qDrop & 0.94 & 0.97 & 1233 & 1158 & \multirow{2}{*}{0.89} & 1158 & 0 & 0 \\ \cline{2-6} \cline{8-10}
& wDrop & 0.75 & 0.70 & 251 & 292 &  & 75 & 189 & 28 \\ \hline

\textbf{Gradient Boosting Classifier} & qDrop & 0.72 & 0.83 & 1233 & 883 & \multirow{2}{*}{0.84} & 883 & 0 & 0 \\ \cline{2-6} \cline{8-10}
& wDrop & 0.63 & 0.68 & 251 & 211 &  & 48 & 158 & 5 \\ \hline

\textbf{Logistic Reg Classifier} & qDrop & 0.95 & 0.97 & 1233 & 1184 & \multirow{2}{*}{0.88} & 1174 & 10 & 0 \\ \cline{2-6} \cline{8-10}
& wDrop & 0.62 & 0.68 & 251 & 205 &  & 35 & 155 & 15 \\ \hline

\textbf{Decision Tree Classifier} & qDrop & 0.72 & 0.83 & 1233 & 883 & \multirow{2}{*}{0.77} & 883 & 0 & 0 \\ \cline{2-6} \cline{8-10}
& wDrop & 0.67 & 0.47 & 251 & 472 &  & 273 & 173 & 26 \\ \hline

\end{tabular}
\label{table-2}
\end{table*}

\begin{table*}[ht]
\centering
\caption{Ablation study of all the selected ML models}
\begin{tabular}{|l|c|c|c|c|c|c|c|c|c|c|c|}
\hline
\textbf{Features Removed} & \multicolumn{2}{c|}{\textbf{1. RF }} & \multicolumn{2}{c|}{\textbf{2. KNN}} & \multicolumn{2}{c|}{\textbf{3. GBOOST}} & \multicolumn{2}{c|}{\textbf{4. LR }} & \multicolumn{2}{c|}{\textbf{5. DT }} \\ \cline{2-11}
 & \textbf{Recall} & \textbf{F1} & \textbf{Recall} & \textbf{F1} & \textbf{Recall} & \textbf{F1} & \textbf{Recall} & \textbf{F1} & \textbf{Recall} & \textbf{F1} \\ \hline
\textbf{Jitter} & 0.85 & 0.88 & 0.87 & 0.89 & 0.82 & 0.86 & 0.79 & 0.85 & 0.80 & 0.77 \\ \hline
\textbf{RTT} & 0.82 & 0.68 & 0.86 & 0.76 & 0.79 & 0.85 & 0.79 & 0.84 & 0.82 & 0.68 \\ \hline
\textbf{cWnd} & 0.41 & 0.44 & 0.46 & 0.50 & 0.39 & 0.42 & 0.74 & 0.79 & 0.43 & 0.45 \\ \hline
\textbf{Jitter \& RTT} & 0.48 & 0.53 & 0.46 & 0.53 & 0.48 & 0.52 & 0.76 & 0.81 & 0.52 & 0.39 \\ \hline
\textbf{cWnd, Jitter, RTT} & 0.37 & 0.33 & 0.37 & 0.33 & 0.33 & 0.33 & 0.32 & 0.31 & 0.37 & 0.33 \\ \hline
\textbf{All included} & 0.85 & 0.88 & 0.90 & 0.89 & 0.78 & 0.84 & 0.86 & 0.88 & 0.79 & 0.77 \\ \hline
\end{tabular}
\label{table-3}
\end{table*}

\section{Results}
In this section, we present the performance evaluation of the selected classifiers: RF, KNN, GB, LR, and DT, based on their ability to classify packets into two types: qDrop, and wDrop. The metrics used for evaluation include Recall, F1-Score, Support, Macro Average F1, and the Confusion Matrix. The detailed results are summarized in Table \ref{table-2}. \\

The evaluation of these five classifiers on packet loss classification revealed key performance differences based on Recall, F1-Score, and the Confusion Matrix. The RF and KNN classifiers both achieved high recall and F1-scores for qDrop, indicating their strong ability to correctly identify qDrop instances. However, their performance drops for wDrop, with lower recall and F1-scores, reflecting challenges in accurately predicting this packet type. LR performed similarly well for qDrop but slightly better for wDrop compared to RF, as evidenced by its higher F1-score for wDrop. GB showed moderate performance, with a noticeable decrease in recall and F1-score for both packet types, particularly for qDrop, indicating that it struggled more with accurate classification in this dataset. The DT classifier demonstrated the weakest performance overall, particularly for wDrop, where it had the lowest F1-score among the models.

The Macro Average F1 scores, which average the F1-scores across both classes, indicate that RF and KNN classifiers are the most balanced across packet types, with the highest scores of $0.88$ and $0.89$, respectively. The Confusion Matrix highlights the specific areas of misclassification for each model, with RF, LR, and KNN showing fewer misclassifications compared to the other models.  Overall, the RF and KNN models emerged as the most effective classifiers for this task, with better overall performance in classifying packet loss types in the network traffic data.

\subsection{Ablation study}

Table \ref{table-3} presents the results of an ablation study conducted on the selected machine learning models to assess how the removal of specific features impacts the models' performance in predicting packet loss types. The features removed include jitter, RTT, cWnd, and combinations thereof. The metrics evaluated are macro average Recall and F1-Score, which reflect the models' ability to correctly identify packet loss instances and the balance between precision and recall.

\subsubsection{Single Feature Removal}
Removing jitter causes a modest decrease in performance across all models. KNN and RF maintain relatively high recall and F1-scores, indicating robustness even without this feature. GB and LR also perform well, though slightly less effectively. The removal of RTT has a more pronounced effect, particularly on the RF, GB, and DT models, all of which show a significant drop in F1-scores. KNN is the least affected, maintaining an F1-score of $0.76$. Removing the cWnd feature results in the most considerable drop in performance among the single feature removals, with all models showing marked decreases in both Recall and F1-Score. RF and KNN show the most resilience, though their performance still diminishes significantly.

\subsubsection{Multiple Feature Removal}
The combined removal of jitter and RTT further degrades performance across all models, with GB and DT experiencing the most significant drops in F1-score, highlighting their reliance on these features for accurate prediction.
The removal of cWnd, jitter, and RTT together results in the lowest performance across the board. F1-scores for all models drop significantly, with most models scoring in the low $0.30s$. This indicates that these features are crucial for the effective functioning of all models.

This ablation study reveals the relative importance of various features for the selected machine learning models in predicting packet loss types. Jitter and RTT emerge as particularly important features, as their removal causes a notable decline in model performance, especially for RF, GB, and DT. The cWnd feature also plays a critical role, with its absence leading to the most substantial performance drops. The study demonstrates that KNN and RF are generally more resilient to feature removal, maintaining higher recall and F1-scores across different scenarios compared to other models. On the other hand, LR and DT tend to perform less effectively when key features are missing, indicating their higher dependency on these features for accurate predictions. The overall findings underscore the importance of including these critical features to ensure high prediction accuracy in packet loss classification tasks.

%\begin{figure}[h]
%\centering
%\includegraphics[width=2.5in]{figures/diagrams.pdf}
%\caption{Network Scneario}
%\label{fig_net}
%\end{figure}

\section{Conclusion}
In this paper, we explored the application of machine learning models to predict and differentiate between congestive and non-congestive packet losses in a simulated network. Using the ns-3 simulator, we trained and evaluated five classifiers: Random Forest, K-Nearest Neighbor, Gradient Boosting, Logistic Regression, and Decision Tree. This study demonstrated the effectiveness of machine learning models, particularly Random Forest and K-Nearest Neighbor, in accurately predicting and differentiating between congestive and non-congestive packet losses. Through extensive evaluation and an ablation study, we identified the critical role of features like jitter, RTT, and cWnd in ensuring high model performance. Random Forest and K-Nearest Neighbor consistently outperformed other models, maintaining high recall and F1-scores, highlighting their robustness and reliability. These findings suggest that integrating these models into live networks could significantly enhance network performance. Our future work will focus on refining these models for practical deployment in dynamic network environments.

\ifCLASSOPTIONcompsoc
  \section*{Acknowledgments}
\else
  \section*{Acknowledgment}
\fi

This work was supported by the Electronics and Telecommunications Research Institute (ETRI) grant funded by the ICT  R\&D program of MSIT/IITP[2021-0-00715, Development of  End-to-End Ultra-high Precision Network Technologies ].



%\bibliographystyle{IEEEtran}

\end{document}